\documentclass{JHEP3}

\usepackage{graphicx}
\usepackage{latexsym}
\usepackage{amsmath}
\usepackage{amsfonts}
\usepackage{amssymb}

\newcommand{\g}{\gamma}

\newcommand{\bb}{\bibitem}

\def\bb{\bibitem}
\def\as{A\!\!\!/}
\def\ps{p\!\!\!/}
\def\bs{b\!\!\!/}

\def\g{\gamma}

\def\bb{\bibitem}
\newcommand{\be}{\begin{equation}}
\newcommand{\ee}{\end{equation}}
\newcommand{\bea}{\begin{eqnarray}}
\newcommand{\eea}{\end{eqnarray}}

\newcommand{\ben}{\begin{eqnarray}}
\newcommand{\een}{\end{eqnarray}}
\newcommand{\bes}{\begin{subequations}}
\newcommand{\ees}{\end{subequations}}

\title{Lorentz-CPT violation, radiative corrections and finite temperature}

\author{Jose R. Nascimento,$^{a,b}$ Eduardo Passos,$^{a}$ and Albert Yu. Petrov$^{a}$ \\
$^{a}$Departamento de F\'\i sica, Universidade Federal da Para\'\i
ba, Caixa Postal 5008, 58051-970 Jo\~ao Pessoa, Para\'\i ba,
Brazil\\
$^{b}$Instituto de F\'\i sica, Universidade de S\~ao Paulo, Caixa
Postal 66318, 05315-970, S\~ao Paulo, SP,
Brazil\\
E-mail:{ jroberto@fisica.ufpb.br, passos@fisica.ufpb.br,
petrov@fisica.ufpb.br, jroberto@fma.if.usp.br}}

\author{Francisco A. Brito \\
Departamento de F\'\i sica, Universidade Federal de Campina Grande,
Caixa Postal 10071, 58109-970  Campina Grande, Para\'\i ba,
Brazil\\
E-mail:{ fabrito@df.ufcg.edu.br}}

\abstract{In this work we investigate the radiatively induced
Chern-Simons-like terms in four-dimensions at zero and finite
temperature. We use the approach of rationalizing the fermion
propagator up to the leading order in the CPT-violating coupling
$b_\mu$. In this approach, we have shown that although the
coefficient of Chern-Simons term can be found unambiguously in
different regularization schemes at zero or finite temperature, it
remains undetermined. We observe a correspondence among results
obtained at finite and zero temperature.
 \\

\vspace{1cm}

Keywords: Chern-Simons Theories, Space-Time Symmetries, Thermal
Field Theory}

\maketitle

\begin{document}

%%%%%%%%%%%%%%%%%%%%%%%%%%%%%
\section{Introduction}
%%%%%%%%%%%%%%%%%%%%%%%%%%%%%

In recent years, it has been investigated in the literature the
possibility of the Lorentz and CPT symmetries being violated in the
nature \cite{0,1,2,3,4,5,6,7,8,9,10,11,11.2,12,13,14}. Theoretical
investigations have pointed out that these symmetries can be
approximate. The realization of this violation can be obtained in
QED by adding the Chern-Simons-like term with a constant
quadrivector  $k_\mu$,
$\frac{1}{2}k_\mu\epsilon^{\mu\alpha\beta\gamma}F_{\alpha\beta}A_\gamma$,
to the Maxwell's theory, and another term which is a CPT-odd term
for fermions, i.e., $\bar{\psi}{\bs}\gamma_5\psi$ with a constant
quadrivector $b_\mu$. This extension of the QED does not break the
gauge symmetry of the action and equations of motion but it does
modify the dispersion relations for different polarization of
photons and Dirac's spinors. Interesting investigations in the
context of Lorentz-CPT violation have appeared recently in the
literature, where several issues were addressed, such as
\v{C}erenkov-type mechanism called ``vacuum \v{C}erenkov radiation''
to test the Lorentz symmetry \cite{ptl}, changing of gravitational
redshifts for differently polarized Maxwell-Chern-Simons photons
\cite{kkl}, evidence for the Lorentz-CPT violation from the
measurement of CMB polarization \cite{bjx}, supersymmetric
extensions \cite{bch}, breaking of the Lorentz group down to the
little group associated with $k_\mu$ \cite{hle} and magnetic
monopoles inducing electric current \cite{brz}.

The dynamical origin of the parameters $k_\mu$ and $b_\mu$ present
in the Lorentz and CPT symmetry breaking is obtained when we
integrate over the fermion fields in the modified Dirac action such
that the radiative corrections may lead to $k_\mu=C b_\mu$. Several
studies have shown that $C$ can be found to be finite but
undetermined \cite{15,16,17,18,Bonneau}.

In the present work, we focus attention on induced Chern-Simons-like
terms via radiative corrections both in zero and finite temperature
by using different regularization schemes. Our starting point here
is the use of the approach of rationalizing the fermion propagator
up to the leading order in the CPT-violating coupling $b_\mu$, after
using the derivative expansion method. By using this approach we
have obtained new and previous results of the literature and also
observe the existence of new effects at finite temperature. Then, by
using dimensional regularization and Lorentz invariant
regularization schemes at zero temperature, we show that the
Chern-Simons coefficient is found to be unambiguously finite, but
with different values. We found that these coefficients are indeed
limits of the theory when we take into account finite temperature
and only use the Lorentz invariant regularization and derivative
expansion method. On the other hand, by using dimensional
regularization at finite temperature, in our present approach, leads
to other finite results in zero and finite temperature found in the
literature.

\section{Radiative corrections }
\label{one}
%%%%%%%%%%%%%%%%%%%%%%%%%%%%%
The one-loop effective action  $S_{\rm eff}[b,A(x)]$ of the gauge
field $A(x)$ can be expressed in the form of the following
functional trace \be S_{\rm eff}[b,A(x)]=-i\,{\rm Tr}\,\ln(\ps- m -
\bs\gamma_5 -e\as(x) ). \ee This functional trace can be represented
as $S_{\rm eff}[b,A(x)]=S_{\rm eff}[b]+S_{\rm
eff}^{\,\prime}[b,A(x)]$. The first term $S_{\rm eff}[b]=-i\,{\rm
Tr}\ln(\ps- m - \bs \gamma_5 )$ does not depend on the gauge field,
and the only nontrivial dynamics is concentrated in the second term
$S_{\rm eff}^{\,\prime}[b,A(x)]$, which is given by the following
power series \be\label{ea} S_{\rm eff}^{\,\prime}[b,A(x)]=i\,{\rm
Tr} \sum_{n=1}^{\infty}\frac1n \Biggl[\frac1{\ps- m - \bs \gamma_5
}\,e \as(x)\Biggr]^n. \ee To obtain the Chern-Simons term we should
expand this expression up to the second order in the gauge field \be
\label{Ef1} S_{\rm eff}^{\,\prime}[b,A(x)]=S_{\rm
eff}^{(2)}[b,A(x)]+\ldots, \ee where \be S_{\rm
eff}^{(2)}[b,A(x)]=\frac{-ie^{2}}{2}{\rm
Tr}\bigl[S_{b}(p)\;\as(x)\;S_{b}(p)\;\as(x)\;\bigl], \ee with
$S_{b}(p)$ being the exact fermion propagator expressed in the form
\be\label{pf} S_{b}(p)=\frac{i}{\ps- m - \bs\gamma_5}. \ee Using the
derivative expansion method \cite{de} one can find that the first
one-loop contribution to $S_{\rm eff}^{(2)}[b,A]$ reads
\be\label{Ef3} S_{\rm eff}^{(2)}[b,A(x)]=\frac{1}{2}\int
d^4x\;\Pi^{\alpha\mu\nu}\partial_{\alpha}A_{\mu}A_{\nu}, \ee where
the one-loop self-energy $\Pi^{\alpha\mu\nu}$ is given by
\be\label{I1}
\Pi^{\alpha\mu\nu}=-ie^{2}\int\,\frac{d^{4}p}{(2\pi)^{4}}\,\text{tr}\bigl[S_{b}(p)\,\gamma^
{\mu}\,S_{b}(p)\,\gamma^{\alpha}\,S_{b}(p)\,\gamma^{\nu}\bigl]. \ee

%\subsection{Approach one}
We now focus on the study of possible indetermination of the
Chern-Simons coefficient that can appear when we expand the
self-energy (\ref{I1}) as in the approach of rationalizing the
fermion propagator up to the leading order in $b$. Then, we may use
the approximation scheme developed in \cite{Ebert} to expand the
exact propagator in Eq.(\ref{pf}) up to the first order in $b$
\be\label{fat}
S_{b}(p)=i\Big[\frac{\ps+m-\g_{5}\bs}{(p^{2}-m^{2})}-\frac{2\g_{5}(m\bs-(b\cdot
p))(\ps+m)}{(p^{2}-m^{2})^{2}}\Big]+\cdots, \ee where we have
retained only the leading order terms in  $b$. Before everything,
let us use the following notation \be\label{sipl1}
S_{b}(p)=i(W_{1}+W_{2}\g_{5}), \ee  where \bea\label{prop12}
W_{1}&=&\frac{\ps+m}{(p^{2}-m^{2})},\\
\label{prop12.2}W_{2}&=&\frac{\bs}{(p^{2}-m^{2})}-\frac{\bigl[2(m
\bs+(b\cdot p))(\ps-m)\bigl]}{(p^{2}-m^{2})^{2}}, \eea into
Eq.(\ref{I1}) to get the form \bea\label{I2}
\Pi^{\mu\alpha\nu}=&&-e^{2}\int\frac{d^{4}p}{(2\pi)^{4}}
\text{tr}\bigl[W_{1}\g^{\mu}W_{1}\g^{\alpha}W_{2}\g_{5}\g^{\nu}
+W_{1}\g^{\mu}W_{2}\g_{5}\g^{\alpha}W_{1}\g^{\nu}\nonumber\\
&&+W_{2}\g_{5}\g^{\mu}W_{1}\g^{\alpha}W_{1}\g^{\nu}\bigl]. \eea
Substituting (\ref{prop12}) and (\ref{prop12.2}) into (\ref{I2}) and
using the relation $\{\g^{\mu}, \g_{5}\}=0$, we can calculate the
trace of gamma matrices, resulting in the following expression
\bea\label{D2}
\Pi^{\mu\alpha\nu}=&&4ie^{2}\int\,\frac{d^{4}p}{(2\pi)^{4}}\frac{1}{(p^{2}-m^{2})^{3}}
\{3\varepsilon^{\alpha\mu\nu\theta}\bigl[b_{\theta}(p^{2}+m^{2})-2p_{\theta}(b\cdot
p)\bigl]+\nonumber\\&-&2b_{\theta}\bigl[\varepsilon^{\beta\mu\nu\theta}p_{\beta}p^{\alpha}
+\varepsilon^{\alpha\beta\nu\theta}p_{\beta}p^{\mu}+\varepsilon^{\alpha\mu\beta\theta}
p_{\beta}p^{\nu}\bigl]\}. \eea Note that by power counting, the
momentum integral in (\ref{D2}) involves finite terms and terms with
logarithmic divergence. Because the integral is divergent, there is
no unique answer regardless of the regularization scheme used, e.g.,
Pauli-Villars or dimensional regularization \cite{15}. Here, we
shall adopt simpler regularization schemes in order to calculate the
divergent integral, such as dimensional regularization \cite{GM} and
Lorentz preserving regularization. The above integral is promoted to
$D$ dimensions and a straightforward calculation yields \be
\Pi^{\mu\alpha\nu}=\frac{6e^{2}}{(4\pi)^{D/2}}\,\frac{\epsilon\,\Gamma(\epsilon/2)}{(m^{2})^{\epsilon/2}}\,\varepsilon^{\mu\alpha\nu\theta}
\,b_{\theta}, \ee where $\epsilon=4-D$. Therefore, for $D=4$, we
find \be\label{Ef6} S_{\rm
eff}^{(2.2)}[b,A(x)]=\frac{3e^{2}}{8\pi^{2}}\int
d^4x\;b_{\beta}\varepsilon^{\mu\alpha\nu\beta}\partial_{\alpha}A_{\mu}A_{\nu}.
\ee This shows that the Chern-Simons coefficient $k_{\beta}$ relates
with $b_{\beta}$ is the form
\ben\label{c1}k_{\beta}=\frac{3e^{2}}{8\pi^{2}}b_{\beta}.\een On the
other hand, there also exists the possibility of using in
Eq.(\ref{D2}) the relation \be\label{lorentz}
\int\frac{d^{4}p}{(2\pi)^{4}}p_{\mu}p_{\nu}\,f(p^{2})=
\frac{g_{\mu\nu}}{4}\int\frac{d^{4}p}{(2\pi)^{4}}p^{2}\,f(p^{2}),
\ee that naturally removes the logarithmic divergence. As a result,
we have only the finite contribution \be
\Pi^{\mu\alpha\nu}=\frac{6e^{2}}{(4\pi)^{D/2}}\,\frac{\Gamma(1+\epsilon/2)}
{(m^{2})^{\epsilon/2}}\,\varepsilon^{\mu\alpha\nu\theta}
\,b_{\theta}. \ee Therefore, for $D=4$, we find \be\label{Ef7}
S_{\rm eff}^{(2.2)}[b,A(x)]=\frac{3e^{2}}{16\pi^{2}}\int
d^4x\;b_{\beta}\varepsilon^{\mu\alpha\nu\beta}\partial_{\alpha}A_{\mu}A_{\nu},
\ee which shows that the Chern-Simons coefficient $k_{\beta}$ now
relates with $b_{\beta}$ in the form \ben \label{c2}
k_{\beta}=\frac{3e^{2}}{16\pi^{2}}b_{\beta}.\een

Although the results (\ref{c1}) and (\ref{c2}) were obtained
unambiguously in different regularization schemes, they show that
$k_\beta$ is indeed undetermined. One the other hand, it is
interesting to note that these results were shown to be connected to
each other at finite temperature \cite{N3}. The results (\ref{c1})
and (\ref{c2}) are limits of high temperature and zero temperature,
respectively, as it was obtained in \cite{N3} where the logarithmic
divergences were eliminated by using the Lorentz preserving
regularization (\ref{lorentz}).

\section{Finite temperature effects}
\label{two}
%%%%%%%%%%%%%%%%%%%%%%%%%%%%%%%%%%%%%%%%%

Let us now study the Chern-Simons coefficient when we take
temperature into account. The effect of high temperature in the
context of breaking Lorentz and CPT symmetries has generated
interesting studies in the literature
\cite{Ebert,N3,Ze1,grig,pass7}. However, only \cite{Ebert} uses the
approach (\ref{fat}) of  rationalizing the fermion propagator up to
the leading order in $b$ in the context of finite temperature.
Following such an approach, in this section, we are going to
investigate the Chern-Simons coefficient at finite temperature.

To develop calculations with finite temperature, let us now assume
that the system is in the state of the thermal equilibrium with a
temperature $T=1/{\beta}$. In this case we can use the Matsubara
formalism for fermions. This consists of taking $p_{0}\equiv
\omega_n=(n+1/2)\frac{2\pi}{\beta}$ and replacing the integration
over zeroth component of the momentum by a discrete sum
$(1/2\pi)\int dp_{0}\rightarrow\frac{1}{\beta}\sum_{n}$
\cite{Dolan}. We also change the Minkowski space to Euclidean space
by performing the Wick rotation $x_{0}\to-ix_{0}$, $p_{0}\to
ip_{0}$, $b_{0}\to ib_{0}$, $d^{4}x\to -id^{4}x$ and $d^{4}p\to
id^{4}p$.
%%%%%%%%%%%%%%%%%%%%%%%%%%%%%%%%%%%%%%%%%%%%%%%%%%%%%%%%%%%%%%%%%%%%%%%%%%%%%%%%%%
%\subsection{Approach two}
Considering initially the expression (\ref{D2}), we have that the
effect of the temperature must reproduce the following structure
\bea \Pi^{\mu\alpha\nu}=&&\frac{-4 i
e^{2}}{\beta}\sum^{\infty}_{n=-\infty}\int\,\frac{d^{3}\vec{p}}{(2\pi)^{3}}\,\frac{1}{(\vec{p}^{\,2}
+M_{n}^{2})^{3}}\{3\varepsilon^{\alpha\mu\nu\theta}\bigl[b_{\theta}(\vec{p}^{\,2}+M_{n}^{2}-2m^{2})-2p_{\theta}(b\cdot
p)\bigl]+\nonumber\\&-&2b_{\theta}\bigl[\varepsilon^{\beta\mu\nu\theta}p_{\beta}p^{\alpha}
+\varepsilon^{\alpha\beta\nu\theta}p_{\beta}p^{\mu}+\varepsilon^{\alpha\mu\beta\theta}p_{\beta}p^{\nu}\bigl]\},
\eea where \bea M^2_n=(n+\frac{1}{2})^2\frac{4\pi^2}{\beta^2}+m^2.
\eea To implement translation only on the space coordinates of the
loop momentum $p_{\rho}$ we decompose it as follows \cite{grig}\bea
p_{\rho}\to\vec{p}_{\rho}+p_0\delta_{0\rho}. \eea We use the
covariance under spatial rotations which allows us to carry out the
following replacement \bea
\vec{p}_{\rho}\vec{p}^{\sigma}\to\frac{\vec{p}^2}{D}(\delta^{\sigma}_{\rho}-\delta_{\rho
0}\delta^{\sigma}_0). \eea Thus, \bea
&&2p_{\rho}b_{\sigma}p^{\sigma}\to
2\left(b_{\rho}\frac{\vec{p}^2}{D}-b_0\delta_{\rho
0}(\frac{\vec{p}^2}{D}-p^2_0)
\right),\nonumber\\&&2p_{\beta}p^{\alpha}\to2\left(\delta_{\beta}^{\alpha}\,\frac{\vec{p}^{\,2}}{D}-\delta_{\beta
0}\delta_{0}^{\alpha}(\frac{\vec{p}^2}{D}-p^2_0)\right),\nonumber\\&&2p_{\beta}p^{\mu}\to
2\left(\delta_{\beta}^{\mu}\,\frac{\vec{p}^{\,2}}{D}-\delta_{\beta
0}\delta_{0}^{\mu}(\frac{\vec{p}^2}{D}-p^2_0)\right),\nonumber\\&&2p_{\beta}p^{\nu}\to
2\left(\delta_{\beta}^{\nu}\,\frac{\vec{p}^{\,2}}{D}-\delta_{\beta
0}\delta_{0}^{\nu}(\frac{\vec{p}^2}{D}-p^2_0)\right). \eea We have
that only the terms above can contribute to the Chern-Simons
structure. Therefore, we can split the expression (\ref{Ef3}) into a
sum of two parts, ``covariant" and ``noncovariant", i.e., \bea
S^{\rm cov}_{\rm
eff}=(-i)\int\,d^{4}x\,I_{1}(m,\beta)\varepsilon^{\alpha\mu\nu\beta}\,b_{\beta}
\partial_{\alpha}A_{\mu}A_{\nu},
\eea with \bea
I_{1}(m,\beta)=\frac{-6ie^{2}}{\beta}\,\sum^{\infty}_{n=-\infty}\int\,\frac{d^D\vec{p}}{(2\pi)^D}
\,\frac{(1-\frac{4}{D})\vec{p}^{2}+M^2_n-2m^{2}}{(\vec{p}^{2}+M^{2}_n)^{3}},
\eea and \bea S^{\rm ncv}_{\rm
eff}&=&i\int\,d^{4}xI_{2}(m,\beta)\,\bigl[3\varepsilon^{\alpha\mu\nu
0}\,b_{0}\partial_{\alpha}A_{\mu}A_{\nu}+\nonumber\\&+&
b_{\theta}(\varepsilon^{0\mu\nu\theta}\partial_{0}A_{\mu}A_{\nu}+\varepsilon^{\alpha
0 \nu\theta}\partial_{\alpha}A_{0}A_{\nu}+\varepsilon^{\alpha\mu
0\theta}\partial_{\alpha}A_{\mu}A_{0})\bigl], \eea with \bea
I_{2}(m,\beta)=\frac{4ie^{2}}{\beta}\,\sum^{\infty}_{n=-\infty}\int\,\frac{d^D\vec{p}}{(2\pi)^D}
\,\frac{\frac{\vec{p}^{2}}{D}-M^2_n+m^{2}}{(\vec{p}^{2}+M^{2}_n)^{3}}.
\eea After integration over the spatial momentum, we find \bea
I_{1}(m,\beta)&=&\frac{-3ie^{2}}{(4\pi)^{D/2}\beta}\,\sum^{\infty}_{n=-\infty}
\left[\frac{(\Gamma(3-\frac{D}{2})-(2-\frac{D}{2})\Gamma(2-\frac{D}{2}))}{(M_{n}^{2})^{2-\frac{D}{2}}}
-\frac{2m^{2}\Gamma(3-\frac{D}{2})}{(M_{n}^{2})^{3-\frac{D}{2}}}\right],
\nonumber\\&=&\frac{6im^{2}e^{2}\Gamma(\lambda_{2})}{(4\pi)^{D/2}\beta}
\,\left(\frac{a^{2}}{m^{2}}\right)^{\lambda_{2}}\sum^{\infty}_{n=-\infty}
\,\frac{1}{((n+b)^{2}+a^{2})^{ \lambda_{2}}}, \eea and \bea
I_{2}(m,\beta)&=&\frac{ie^{2}}{(4\pi)^{D/2}\beta}\sum^{\infty}_{n=-\infty}
\left[\frac{\epsilon^{\prime}\Gamma(2-\frac{D}{2})}{(M_{n}^{2})^{2-\frac{D}{2}}}
-\frac{2m^{2}\Gamma(3-\frac{D}{2})}{(M_{n}^{2})^{3-\frac{D}{2}}}\right],
\nonumber\\&\!=\!&\frac{ie^{2}}{(4\pi)^{D/2}\beta}\!\!\sum^{\infty}_{n=-\infty}\!\!
\left[\left(\!\frac{a^{2}}{m^{2}}\!\right)^{\lambda_{1}}\!\!\!\frac{\epsilon^{\prime}
\Gamma(\lambda_{1})}{((n+b)^{2}+a^{2})^{\lambda_{1}}}\!-\!\left(\!\frac{a^{2}}{m^{2}}\!\right)^{\lambda_{2}}
\!\!\!\frac{2m^{2}\Gamma(\lambda_{2})}{((n+b)^{2}+a^{2})^{\lambda_{2}}}\right]\mbox{
} \eea where $\lambda_1=2-\frac{D}{2}$, $\lambda_2=3-\frac{D}{2}$,
$\epsilon^{\prime}=3-D$, $a=m\beta/2\pi$ and $b=1/2$.

\begin{figure}[h]
\centerline{\includegraphics[{angle=90,height=7.0cm,angle=180,width=8.0cm}]
{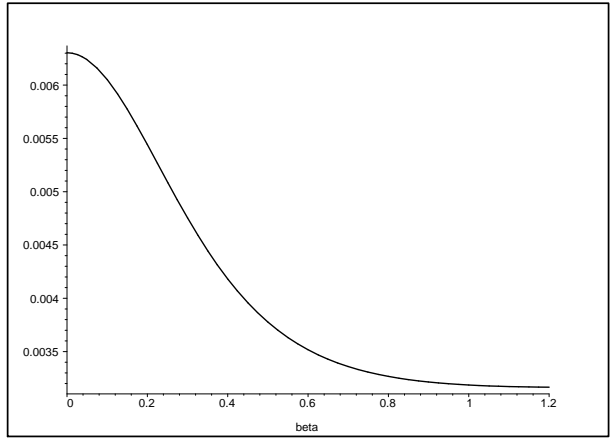}} \caption{The function $F(a)$ is diferent from zero
everywhere. At zero temperature ($\beta\to\infty$), the function
tends to a nonzero value.}\label{fig1}
\end{figure}

At this point we need an explicit expression for the sum over the
Matsubara frequencies. We use the following result \cite{FO}
\bea\label{fo} \sum_n \frac{1}{[(n+b)^2 + a^2]^{\lambda}}&=&
\frac{\sqrt{\pi}\Gamma(\lambda - 1/2)}{\Gamma(\lambda)(a^2)^{\lambda
- 1/2}}
+\nonumber\\
&+&4\sin(\pi\lambda)\int_{|a|}^\infty \frac{dz}{(z^2 -
a^2)^{\lambda}} Re\left(\frac{1}{\exp 2\pi(z + ib) -1}\right), \eea
which is valid for $1/2<\lambda<1$. Note that for
$\lambda_1=2-\frac{D}{2}$ and $\lambda_2=3-\frac{D}{2}$ we cannot
apply this relation for $D=3$ since the integral diverges. Thus, we
carry out the analytic continuation of this relation, so that we
obtain \bea \label{ancont} &&\int_{|a|}^\infty \frac{dz}{(z^2 -
a^2)^{\lambda}}
Re\left(\frac{1}{\exp 2\pi(z + ib) -1}\right)= \nonumber \\
&=& \frac{1}{2a^2}\frac{3-2\lambda}{1-\lambda} \int_{|a|}^\infty
\frac{dz}{(z^2 - a^2)^{\lambda-1}} Re\left(\frac{1}{\exp 2\pi(z +
ib) -1}\right)\\  &-& \frac{1}{4a^2}\frac{1}{(2-\lambda)(1-\lambda)}
\int_{|a|}^\infty \frac{dz}{(z^2 - a^2)^{\lambda-2}}
\frac{d^2}{dz^2}Re\left(\frac{1}{\exp 2\pi(z + ib)
-1}\right).\nonumber \eea Now we can substitute this expression into
(\ref{fo}), for $D=3$, so that after some simplifications we get

\be I_{1}(\beta)=\frac{3ie^{2}}{8\pi^{2}}[1+2\pi^{2}F(a)], \ee and

\be I_{2}(\beta)=-\frac{ie^{2}}{4}\,F(a), \ee where the function
$F(a)$ given by \be \label{fp4}
F(a)=\int_{|a|}^{\infty}dz(z^2-a^2)^{1/2} \frac{\tanh(\pi
z)}{\cosh^2(\pi z)}, \ee has the following limits:
$F(a\to\infty)\to0$ ($T\to0$) and $F(a\to0)\to 1/2\pi^{2}$
($T\to\infty$) --- see Fig.\ref{fig1}.

In  summary we find that the Chern-Simons coefficients at finite
temperature for both covariant and noncovariant parts are:

\ben &&k^{\rm cov}_\alpha=\frac{3e^2}{4\pi^2}b_\alpha,\qquad k^{\rm
ncv}_0=\frac{3e^2}{8\pi^2}b_0,\qquad k^{\rm
ncv}_i=\frac{e^2}{8\pi^2}b_i,\qquad \!\!\!\! (a\to0 \mbox{
or } T\to\infty)\\
&&k^{\rm cov}_\alpha=\frac{3e^2}{8\pi^2}b_\alpha,\qquad k^{\rm
ncv}_\alpha=0.\qquad\qquad\qquad\qquad\qquad\qquad (a\to\infty\mbox{
or } T\to0) \een Note that the covariant coefficient
$\frac{3e^2}{8\pi^2}b_\alpha$ at $T\to0$ coincides with the
coefficient previously obtained  at zero temperature (\ref{c1}), and
corresponds to the result obtained in \cite{N3} at the limit
$T\to\infty$. The noncovariant result $\frac{e^2}{8\pi^2}b_i$ at
$T\to\infty$ corresponds to the result found in \cite{11.2,chan} at
zero temperature, and the noncovariant result $k^{\rm ncv}_\alpha=0$
at $T\to0$ corresponds to the result found in \cite{grig} at the
same limit.

%%%%%%%%%%%%%%%%%%%%%%%%%%%%%%%%%%%%%%%%%%%%%%%%%%%%%%%%%%%%%%%%%%%%%%%%%%%%%%%%%%%%%%%%%%%%%%%%%%%%%%%%%%%%

\section{Conclusions}
\label{conclu}

In this work we have used the approach of rationalizing the fermion
propagator up to the leading order in the CPT-violating coupling
$b_\mu$, after using the derivative expansion method. We have shown
that although the coefficient of Chern-Simons term can be found
unambiguously in different regularization schemes at zero and finite
temperature, it remains undetermined. An interesting point, however,
we noted here is the possibility of the temperature making the
connection among the coefficients obtained in different
regularization schemes. In a certain sense, each value found in a
particular scheme seems to correspond to a different scale of the
theory, which can be achieved by increasing or decreasing the
ambient temperature. For example, a particular coefficient found in
zero temperature by adopting one approach corresponds to the same
coefficient at high temperature by adopting another one. A complete
understanding of these issues will require further investigations.

\acknowledgments

This work was partially supported by Conselho Nacional de
Desenvolvimento Cient\'\i fico e Tecnol\'ogico (CNPq) and Funda\c
c\~ao de Amparo \`a Pesquisa do Estado de S\~ao Paulo (FAPESP). The
work by A. Yu. P. has been supported by CNPq-FAPESQ-PB DCR program,
CNPq project No. 350400/2005-9.

%%%%%%%%%%%%%%%%%%%%%%%%%%%%%%%%%%%%%%%%%%%%%%%%%%%%%%%%%%%%%%%%%%%
%%%%%%%%%%%%%%%%%%%%%%%%%%%%%%%%%%%%%%%%%%%%%%%%%%%%%%%%%%%%%%%%%%%%

\end{document}